\pdfoutput=1
\documentclass[%
 reprint,
 preprintnumbers,
 bibnotes,
 amsmath,amssymb,
 aps,
 prl,
]{revtex4-1}

\usepackage[utf8]{inputenc}
\usepackage{graphicx}
\usepackage{tikz}
\usepackage{dcolumn}
\usepackage{bm}
\usepackage{hyperref}
\hypersetup{
    colorlinks=true,
    linkcolor=black,
    filecolor=black,      
    urlcolor=black,
    citecolor=black,
}

\begin{document}


\title{Two-loop Octagons, Algebraic Letters and $\bar{Q}$ Equations}
\author{Chi Zhang}%
 \email{zhangchi@itp.ac.cn}
\author{Zhenjie Li}
 \email{lizhenjie@itp.ac.cn}
\author{Song He}
\email{songhe@itp.ac.cn\\}
\affiliation{%
CAS Key Laboratory of Theoretical Physics, Institute of Theoretical Physics, Chinese Academy of Sciences, Beijing 100190, China
}%
\affiliation{%
School of Physical Sciences, University of Chinese Academy of Sciences, No.19A Yuquan Road, Beijing 100049, China
}%

\date{\today}

\begin{abstract}
We compute the symbol of the first two-loop amplitudes in planar ${\cal N}=4$ SYM with algebraic letters, the eight-point NMHV amplitude (or the dual octagon Wilson loops). We show how to apply $\bar{Q}$ equations of~\cite{CaronHuot:2011kk} for computing the differential of two-loop $n$-point NMHV amplitudes and present the result for $n=8$ explicitly. The symbol alphabet for octagon consists of 180 independent rational letters and 18 algebraic ones involving Gram-determinant square roots. We comment on all-loop predictions for final entries and aspects of the result valid for all multiplicities. 
\end{abstract}

\maketitle

\section{Introduction}

Scattering amplitudes are central objects in fundamental physics: not only do they connect to high energy experiments such as Large Hadron Collider, but also provide new insights into Quantum Field Theory (QFT) itself. In recent years, tremendous progress has been made in unravelling hidden structures of scattering amplitudes in gauge theories and gravity, especially in planar ${\cal N} = 4$ supersymmetric Yang-Mills theory (SYM). Integrability of the theory (see \cite{Beisert:2010jr} for a review), and especially pentagon operator product expansion~\cite{Basso:2013vsa} allows us to compute in principle amplitudes or equivalently polygonal light-like Wilson loops at finite coupling. New mathematical structures such as positive Grassmannian and the amplituhedron have led to the computation for all-loop integrands~\cite{ArkaniHamed:2010kv} and even the reformulation of the perturbation theory~\cite{ArkaniHamed:2012nw,Arkani-Hamed:2013jha}. Furthermore, ${\cal N}=4$ SYM has become an extremely fruitful playground for new methods of evaluating Feynman loop integrals, which is a subject of enormous interests ({\it c.f.} \cite{Bourjaily:2018aeq,Henn:2018cdp,Herrmann:2019upk} and references there in). 

Relatedly, there has been formidable progress in computing and understanding multi-loop scattering amplitudes themselves, most notably by the hexagon and heptagon bootstrap program~\cite{Dixon:2011pw,*Dixon:2014xca,*Dixon:2014iba,*Drummond:2014ffa,*Dixon:2015iva,Caron-Huot:2016owq}. The first non-trivial amplitude in planar ${\cal N}=4$ SYM, the six-point amplitude (or hexagon) has been determined through seven and six loops for MHV and NMHV cases respectively~\cite{Caron-Huot:2019vjl}, and similarly the seven-point amplitude (or heptagon) has been determined through four loops for these cases respectively~\cite{Dixon:2016nkn,*Drummond:2018caf}. After subtracting infrared divergences which are known to all loops~\cite{Anastasiou:2003kj}, scattering amplitudes of the theory are transcendental functions of cross-ratios, and for MHV and NMHV cases they are expected to contain only generalized polylogarithms of weight $2L$ at $L$ loops~\cite{ArkaniHamed:2012nw}. There exists a Hopf algebra coaction structure~\cite{Goncharov:2010jf,*Duhr:2011zq,*Duhr:2012fh} and of particular interests is the maximally iterated coproduct called the {\it symbol}. A crucial assumption for the hexagon and heptagon bootstrap is that the collection of letters entering the symbol, or the {\it alphabet}, consists of only $9$ and $42$ variables known as cluster coordinates~\cite{Golden:2013xva}. After imposing conditions on physical discontinuities~\cite{Gaiotto:2011dt,Caron-Huot:2016owq}, the space of such functions are remarkably small and certain information from physical limits and symmetries suffices to fix the result.

It is natural to wonder about scattering amplitudes for $n\geq 8$ (octagons {\it etc.}), which are expected to be much more intricate than $n=6,7$ cases. One important difference is that for $n\geq 8$ we need N${}^k$MHV sectors with $k\geq 2$, which are expected to contain elliptic polylogarithms~\cite{Bourjaily:2017bsb} and more complicated functions~\cite{Bourjaily:2018ycu}. Even when restricting to MHV and NMHV, a challenge for both loop integration and bootstrap method is that we quickly lose control of the symbol alphabet. Starting at $n=8$, the cluster algebra becomes infinite-type and it is unclear which letters can appear for $L$-loop N${}^k$MHV amplitudes (the only data so far being the symbol of two-loop $n$-point MHV amplitudes~\cite{CaronHuot:2011ky}). Moreover, unlike previous cases, a new features is the appearance of {\it algebraic letters} which can no longer be written as rational functions of momentum twistors~\cite{Hodges:2009hk} parametrizing the kinematics. It is of great interests already at two loops to understand what letters with square roots appear for amplitudes (and for Feynman integrals). In this Letter, we compute the symbol of two-loop NMHV octagon using $\bar{Q}$ equations~\cite{CaronHuot:2011kk}, which provides the first example of this phenomenon, and initiate the study for higher-point and even higher-loop amplitudes. 

It is well known that scattering amplitudes in planar ${\cal N}=4$ SYM enjoy both superconformal and dual superconformal symmetries~\cite{Drummond:2006rz,Drummond:2008vq,Korchemsky:2010ut}, which close into the infinite-dimensional Yangian symmetry~\cite{Drummond:2009fd}. In~\cite{CaronHuot:2011kk}, {\it exact} all-loop equations obeyed by the S-matrix were derived by determining the quantum corrections to the generators of Yangian symmetry acting on the Bern-Dixon-Smirnov (BDS)-renormalized S-matrix~\cite{Bern:2005iz}. The equations consist of $\bar{Q}$ equations for dual superconformal symmetry, and its parity-conjugate~\cite{CaronHuot:2011kk}. Perturbatively the equations express the derivatives of amplitudes at a given loop order in terms of integrals of lower-loop amplitudes with more legs, and they can be solved to uniquely determine amplitudes. In practice, MHV and NMHV amplitudes are particularly simple since their differential can be solved using $\bar{Q}$ equations alone, which have been used to (re-)derive the complete symbol of two-loop MHV for all $n$, two-loop NMHV heptagon, and three-loop MHV hexagon~\cite{CaronHuot:2011kk}. Since the equations can be essentially derived from the dual Wilson loop picture, these are computations from first principles, which we can exploit to produce higher-point amplitudes currently not accessible from loop integrals or bootstrap~\footnote{For external kinematics in two dimensions, essentially all two-loop NMHV and three-loop MHV amplitudes have been computed using $\bar{Q}$ equations~\cite{Caron-Huot:2013vda}.}.

As a first step, we compute two-loop NMHV octagon which is the first multi-loop amplitude in planar ${\cal N}=4$ SYM involving algebraic letters. We do so by computing its differential, which via $\bar{Q}$ equations are given by one-dimensional integrals over one-loop N${}^2$MHV amplitudes in the collinear limit. As we will review shortly, the differential of $L$-loop MHV and NMHV amplitudes is given by a combination of generalized polylogarithmic functions of weight $(2L{-}1)$, dressed with certain {\it universal} objects that are independent of loop order; the latter are a collection of Yangian invariants times $d\log$ of {\it final entries} of the symbol, known for MHV to all $n$ and NMHV for $n=6,7$~\cite{CaronHuot:2011ky,CaronHuot:2011kk}. We compute these universal NMHV final entries for all $n$~\cite{toapp}, though here we only need the result for $n=8$. The generalized polylogarithm functions accompanying them have to be computed order by order, and we will present the weight-3 functions for two-loop octagons which in turn gives the complete symbol (there is no qualitative difference for higher $n$, which will be reported elsewhere~\cite{toapp}). 

\begin{figure*}[t]
    \begin{center}
        \begin{tikzpicture}[baseline={([yshift=-.5ex]current bounding box.center)}]
            \node[fill=black,circle,draw=black, inner sep=0pt,minimum size=7pt] at (-2,0) {};
            \node[fill=black,circle,draw=black, inner sep=0pt,minimum size=7pt] at (-2,-1) {};
            \node[fill=black,circle,draw=black, inner sep=0pt,minimum size=7pt] at (-1,-1) {};
            \node[fill=black,circle,draw=black, inner sep=0pt,minimum size=7pt] at (-1,0) {};
            \draw[line width=0.4mm] (-2,0) -- (-2,-1) -- (-1,-1) -- (-1,0) -- cycle;
            \draw[line width=0.4mm] (-2.6,-0.8) -- (-2,-1);
            \draw[line width=0.4mm] (-2,-1) -- (-1.8,-1.6);
            \draw[line width=0.4mm] (-1,-1) -- (-0.4,-0.8);
            \draw[line width=0.4mm] (-1,-1) -- (-1.2,-1.6);
            \draw[line width=0.4mm] (-2.6,-0.2) -- (-2,0);
            \draw[line width=0.4mm] (-1.8,0.6) -- (-2,0);
            \draw[line width=0.4mm] (-1.2,0.6) -- (-1,0);
            \draw[line width=0.4mm] (-1,0) -- (-0.4,-0.2);
            \node at (-1.2,0.8) {$a$};
            \node at (0,-0.2) {$b{-}1$};
            \node at (-0.2,-0.8) {$b$};
            \node at (-1.2,-1.8) {$c{-}1$};
            \node at (-1.8,-1.8) {$c$};
            \node at (-3,-0.8) {$d{-}1$};
            \node at (-2.8,-0.2) {$d$};
            \node at (-1.8,0.8) {$a{-}1$};
            \node at (-2.4,0.1) {$\cdot$};
            \node at (-2.3,0.3) {$\cdot$};
            \node at (-2.1,0.4) {$\cdot$};
            \node at (-0.9,-1.4) {$\cdot$};
            \node at (-0.7,-1.3) {$\cdot$};
            \node at (-0.6,-1.1) {$\cdot$};
            \node at (-0.6,0.1) {$\cdot$};
            \node at (-0.7,0.3) {$\cdot$};
            \node at (-0.9,0.4) {$\cdot$};
            \node at (-2.4,-1.1) {$\cdot$};
            \node at (-2.3,-1.3) {$\cdot$};
            \node at (-2.1,-1.4) {$\cdot$};
        \end{tikzpicture}
        $\displaystyle
            \begin{cases}\displaystyle
                x_{ab}^2 := \frac{\langle a{-}1\,a\,b{-}1\,b\rangle}{\langle a{-}1\,a\rangle\langle b{-}1\,b\rangle},\\[12pt]
                u=\displaystyle\frac{x_{ad}^{2}x_{bc}^{2}}{x_{ac}^{2}x_{bd}^{2}},\quad v=\displaystyle\frac{x_{ab}^{2}x_{cd}^{2}}{x_{ac}^{2}x_{bd}^{2}},\\[12pt]
                \Delta_{abcd} =\sqrt{(1-u-v)^{2}-4uv}
            \end{cases}
        $
    \qquad 
    \begin{tikzpicture}[baseline={([yshift=-.5ex]current bounding box.center)}]
        \node[fill=black,circle,draw=black, inner sep=0pt,minimum size=7pt] at (-2,0) {};
        \node[fill=black,circle,draw=black, inner sep=0pt,minimum size=7pt] at (-2,-1) {};
        \node[fill=black,circle,draw=black, inner sep=0pt,minimum size=7pt] at (-1,-1) {};
        \node[fill=black,circle,draw=black, inner sep=0pt,minimum size=7pt] at (-1,0) {};
        \draw[line width=0.4mm] (-2,0) -- (-2,-1) -- (-1,-1) -- (-1,0) -- cycle;
        \draw[line width=0.4mm] (-2.6,-0.8) -- (-2,-1);
        \draw[line width=0.4mm] (-2,-1) -- (-1.8,-1.6);
        \draw[line width=0.4mm] (-1,-1) -- (-0.4,-0.8);
        \draw[line width=0.4mm] (-1,-1) -- (-1.2,-1.6);
        \draw[line width=0.4mm] (-2.6,-0.2) -- (-2,0);
        \draw[line width=0.4mm] (-1.8,0.6) -- (-2,0);
        \draw[line width=0.4mm] (-1.2,0.6) -- (-1,0);
        \draw[line width=0.4mm] (-1,0) -- (-0.4,-0.2);
        \node at (-1.2,0.8) {$1$};
        \node at (-0.2,-0.2) {$2$};
        \node at (-0.2,-0.8) {$3$};
        \node at (-1.2,-1.8) {$4$};
        \node at (-1.8,-1.8) {$5$};
        \node at (-2.8,-0.8) {$6$};
        \node at (-2.8,-0.2) {$7$};
        \node at (-1.8,0.8) {$8$};
    \end{tikzpicture}	
    \end{center}
    \caption{\label{fig:wide}A general four-mass box (left) and a octagon one (right).}
\end{figure*}
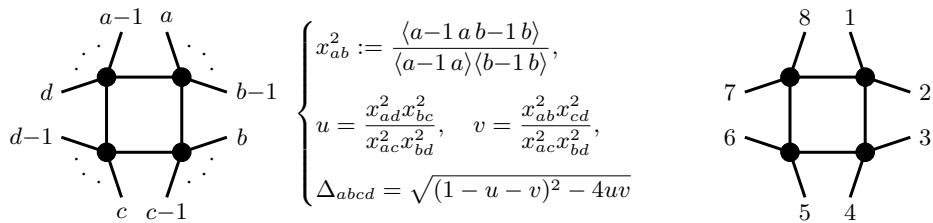

\section{Review of $\bar{Q}$ equations}

The infrared divergences of planar $\mathcal{N}=4$ SYM are captured by the so-called BDS anstaz~\cite{Bern:2005iz}, and we are interested in the infrared-finite object, the \emph{BDS-subtracted amplitude}  $R_{n,k}=A_{n,k}/A_{n}^{\text{BDS}}$,
for $n$-point, N$^{k}$MHV amplitude $A_{n,k}$. As shown in~\cite{CaronHuot:2011kk}, $R_{n,k}$ is dual conformal invariant and enjoys a chiral half of the dual superconformal symmetries; however, it is not invariant under the action of the other half of dual superconformal generators
\begin{equation}
    \bar{Q}_{a}^{A}= \sum_{i=1}^n \chi_{i}^{A}\frac{\partial}{\partial Z_{i}^{a}} \:,
\end{equation}
where $Z_{i}$ are momentum twistors and $\chi_{i}$ denote their Grassmann parts. We further define the basic $\mathrm{SL}(4)$-invariant $\langle ijkl\rangle:=\varepsilon_{abcd}Z_{i}^{a}Z_{j}^{b}Z_{k}^{c}Z_{l}^{d}$ (or the Pl\"{u}cker coordinates of $G(4,n)$), and the basic $R$ invariant~\cite{Drummond:2008vq,Mason:2009qx}
\begin{equation}\label{Rinv}
    [i\,j\,k\,l\,m]:=\frac{\delta^{0\vert 4}(\chi_{i}^{A}\langle jklm\rangle+\text{cyclic})}{\langle ijkl\rangle\langle jklm\rangle
    \langle klmi\rangle\langle lmij\rangle\langle mijk\rangle} \:.
\end{equation}
Nevertheless, this anomaly can be restored by an integral over collinear limits of higher-point amplitudes:
\begin{align}
\bar{Q}_{a}^{A} R_{n,k} = \frac{1}{4}\Gamma_{\rm cusp}~\operatorname{Res}_{\epsilon=0}\int_{\tau=0}^{\tau=\infty}\Bigl(\mathrm{d}^{2\vert 3}\mathcal{Z}_{n+1}\Bigr)_{a}^{A} \nonumber \\
[ R_{n+1,k+1}
-R_{n,k}R_{n+1,1}^{\text{tree}} ]+ \text{cyclic}  \label{Qbar} \:, 
\end{align}
where $\Gamma_{\rm cusp}$ is the cusp anomalous dimension~\cite{Beisert:2006ez}, and we have explicitly shown the term with particle $n{+}1$ added in collinear limit with $n$, and its (super-) momentum twistor $\mathcal{Z}_{n+1}=(Z_{n+1},\chi_{n+1})$ parametrized by $\epsilon, \tau$:
\begin{equation}
    \mathcal{Z}_{n+1}= \mathcal{Z}_{n}- \epsilon \mathcal{Z}_{n-1} + C \epsilon \tau \mathcal{Z}_{1} + C'\epsilon^{2} \mathcal{Z}_{2} \:,
    \label{colpara}
\end{equation}
with $C=\frac{\langle n{-}1\,n\,2\,3\rangle}{\langle n\,1\,2\,3\rangle}$ and $C'=\frac{\langle n{-}2\,n{-}1\,n\,1\rangle}{\langle n{-}2\,n{-1}\,2\,1\rangle}$. The integral measure $\int (\mathrm{d}^{2\vert 3}\mathcal{Z}_{n+1})_{a}^{A}$ consists of the bosonic part $(\mathrm{d}^{2}Z_{n+1})_{a}:=\varepsilon_{abcd}Z_{n+1}^{b}\mathrm{d}Z_{n+1}^{c}\mathrm{d}Z_{n+1}^{d}$ and the fermionic part $(\mathrm{d}^{3}\chi_{n+1})^{A}$; uisng (\ref{colpara}) the bosonic measure reads 
\begin{equation}C(\bar{n})_{a}\operatorname{Res}_{\epsilon=0}\int\epsilon\mathrm{d}\epsilon\int_{0}^{\infty} \mathrm{d}\tau
\end{equation}
with $(\bar{n})_{a}:=(n{-}1\,n\,1)_{a}$. The notation $\operatorname{Res}_{\epsilon=0}$ means to extract the coefficient of $\mathrm{d}\epsilon/\epsilon$ under the collinear limit of $\epsilon\to 0$, and we integrate over ``momentum fraction" $\tau$.
The perturbative expansion of \eqref{Qbar} relates 
$R_{n,k}^{(L)}$ to $ R_{n+1,k+1}^{(L-1)}$ {\it etc.}, and precisely the difference of the two terms in the bracket ensures that the RHS of \eqref{Qbar} is finite. 

For $k=0,1$, the $\bar{Q}$ equations, \eqref{Qbar}, allow us to compute the differential of $R_{n, k}^{(L)}$ which can be expressed as~\footnote{Here it is understood that the differential acts on coefficients of Yangian invariants, but not on Yangian invariants since $\bar{Q} Y_{n,k}=0$.} 
\begin{equation} \label{devofR}
\mathrm{d} R_{n,k}^{(L)}=\sum_\alpha Y_{n,k}^{\alpha}~\mathrm{d}\log(a_\alpha)~{\cal I}^{(2L{-}1)}_\alpha\,.
\end{equation}
Each term consists of (loop-independent) Yangian invariants $Y_{n,k}$ times final entries $\mathrm{d}\log a$, labelled collectively by $\alpha$, multiplied by (pure) transcendental functions of weight $2L{-}1$, ${\cal I}_\alpha$. Note that once we know all functions ${\cal I}^{(2L{-}1)}_\alpha$ (or their symbols), the symbol of $R_{n,k}$ is obtained iteratively as ${\cal S}[ R_{n,k}^{(L)}]=\sum_\alpha Y_{n,k}^{\alpha}~{\cal S} [{\cal I}^{(2L{-}1)}_\alpha] \otimes (a_\alpha)$.

To derive~\eqref{devofR}, first note that all Yangian invariants $Y_{n,k}$ can be computed from positive Grassmannian~ \cite{Drummond:2010uq,ArkaniHamed:2009vw,ArkaniHamed:2012nw} (recall $Y_{n,0}=1$ and all NMHV Yangian invariants are of the form~\eqref{Rinv}), and recursively the RHS of \eqref{Qbar} consists of terms of the form $Y_{n{+}1, k{+}1} F^{(2L{-}2)}_{n+1}$.
A remarkable feature of the integral $\operatorname{Res}_{\epsilon=0}\int \mathrm{d}^{2\vert3}\mathcal{Z}_{n+1}$ is that the fermionic integral $\int\mathrm{d}^{3}\chi_{n+1}$ and residue $\operatorname{Res}_{\epsilon=0}$ part can be performed on Yangian invariants $Y_{n+1,k{+}1}$ independent of transcendental functions $F$'s. One may worry about the $\log^{L-1}\epsilon$ divergences arising from the collinear limit of but the divergences are always canceled after integrating over $\tau$, as shown in~\cite{CaronHuot:2011kk}. For MHV ($k=0$), the effect of $\operatorname{Res}_{\epsilon=0}\int \epsilon\mathrm{d}\epsilon\int \mathrm{d}^{3}\chi_{n+1}$ yields terms of the form 
\begin{equation}
\bar{Q}\log\frac{\langle\bar{n} i\rangle}{\langle\bar{n}j \rangle} \times \int \mathrm{d}\log f_{i,j}(\tau)~F_{n{+}1} (\tau, \epsilon\to 0)
\end{equation}
with some rational function $f_{i,j} (\tau)$ for each term. The last step is trivial for MHV: we can simply replace $\chi_{i}^{A}$ in $\bar{Q}_{a}^{A}$ with $\mathrm{d}Z_{i}^{A}$ then take the trace to obtain the external derivative $\mathrm{d}:=\sum_{i,a}\mathrm{d}Z_{i}^{a}\,\partial/\partial Z_{i}^{a}$, which reproduces the well-known MHV final entries $\mathrm{d}\log \langle i{-}1\, i\, i{+}1\, j\rangle$ after collecting all cyclic terms. The one-dimensional integrals for $F^{(2L{-}2)}_{n{+}1}$ gives weight-$(2L{-}1)$ functions in (\ref{devofR}). 

For NMHV ($k=1$), the effect of $\operatorname{Res}_{\epsilon=0}\int \epsilon\mathrm{d}\epsilon\int \mathrm{d}^{3}\chi_{n+1}$ on N${}^2$MHV Yangian invariants, $Y_{n{+}1,2}$ on the RHS of \eqref{Qbar} gives a list of possible $Y_{n,1}$ in \eqref{Rinv} times final entries as
\begin{equation}\label{NMHVfinal}
Y_{n,1}^\alpha~{\bar Q} \log (a_\alpha) \in \biggl\{  [i\,j\,k\,l\,m]~\bar{Q}\log\frac{\langle\bar{n} I\rangle}{\langle\bar{n} J \rangle} \biggr\}
\end{equation}
where $I, J$ can generally be intersections of momentum twistors of the form {\it e.g.} $(i j) \cap (k l m)$ (see~\cite{ArkaniHamed:2010kv,ArkaniHamed:2010gh}). To obtain the differential of $R_{n,1}$ in this case, the naive replacement above has an ambiguity due to the existence of non-trivial kernel of $\bar{Q}$, which always take the form
\begin{equation}
\bar{Q}_{a}^{A}\biggl([12345]\log\frac{\langle 1234\rangle}{\langle 1235\rangle} \biggr)=0\,. 
       \label{qbarker}      
\end{equation}

Nevertheless, since the kernel of $\bar{Q}$ can not contain non-trivial functions of \emph{dual conformal invariants} (DCI) in this case, the replacement $\chi_{i}\to \mathrm{d}Z_{i}$ has no ambiguity once we convert the arguments of $\bar{Q}\log$ to DCI 
by adding ``0'' of the form~\eqref{qbarker}. It is a straightforward but tedious algorithm to arrive at such a manifestly DCI form, which gives the final answer for $\mathrm{d} R_{n,1}$~\footnote{For N$^{k}$MHV cases with $k\geq 2$, the kernel of $\bar{Q}$ does involve non-trivial dual conformal functions, thus $\bar{Q}$-equation can not determine the result uniquely without supplement with the parity-conjugate, $Q^{(1)}$ equations.}.

\section{\label{sec:3} The Computation}

We start with the chiral box expansion for N${}^2$MHV, BDS-subtracted amplitudes $R_{n{+}1,2}^{(1)}$~\cite{Bourjaily:2013mma}:
\begin{equation}
    R_{n{+}1,2}^{(1)}=\sum_{a<b<c<d}\bigl(f_{a,b,c,d}-R_{n{+}1,2}^{\text{tree}}f_{a,b,c,d}^{\text{MHV}}\bigr)\mathcal{I}_{a,b,c,d}^{\text{fin}}
\end{equation}
where $\mathcal{I}_{a,b,c,d}^{\text{fin}}$ denote the finite part of DCI-regulated box functions which are dilogarithms; the box coefficients $f_{a,b,c,d}$ are one-loop N$^{2}$MHV leading singularities which are linear combinations of $Y_{n{+}1,2}$'s; $f_{a,b,c,d}^{\text{MHV}}=0,1$ are MHV one-loop box coefficients. It is straightforward to perform $\operatorname{Res}_{\epsilon=0}\int \epsilon\mathrm{d}\epsilon \mathrm{d}^{3}\chi_{n+1}$ on $f_{a,b,d,c}$ which gives a list of the form~\eqref{NMHVfinal}, except that a special prescription is needed for four-mass box coefficients $f_{a,b,c,d}$. The latter involve square root of Gram determinant, and they read
\begin{equation*} \label{fourmassbox}
      \frac{1-u-v\pm \Delta}{2\Delta}[\alpha_{\pm},b{-}1,b,c{-}1,c][\gamma_{\pm},d{-}1,d,a{-}1,a] 
\end{equation*}
where $u,v$ and $\Delta$ are shown in Fig. 1, and
$\alpha_{\pm}$ and $\gamma_{\pm}$ are two solutions of the Schubert problem $\alpha= (a{-}1\,a)\cap (d\,d{-1}\,\gamma)$, $\gamma= (c{-}1\,c)\cap (b\,b{-}1\, \alpha)$. It is important to note that in the result produced by such leading singularities, \eqref{NMHVfinal} is accompanied by $\mathrm{d}\log f(\tau)$ where $f(\tau)$ may no longer be a rational function. After $\tau$ integral, they produce algebraic letters in two-loop NMHV amplitudes.

Let's focus on the octagon ($n=8$), where we need to consider nine $9$-point four-mass boxes, and it is clear that only $4$ can potentially contribute square root after taking the collinear limit of $\Delta(\tau)$. They are $f_{2,4,7,9},f_{2,5,7,9},f_{2,4,6,9},f_{3,5,7,9}$ and we see it is clear that in the limit only two square roots can appear, $\Delta_{1,3,5,7}$ (see Fig. 1) and its cyclic image $\Delta_{2,4,6,8}$. The algebraic letters in the answer must involve $z, \bar{z}$ satisfying 
\begin{equation}\label{zzbar}
z\bar{z}=u=\frac{\langle 1278\rangle\langle3456\rangle}{\langle1256\rangle\langle 3478\rangle}\:,\:
        (1-z)(1-\bar{z})=v=\frac{\langle 1234\rangle\langle5678\rangle}{\langle1256\rangle\langle 3478\rangle}\,,
        \end{equation}
and $z', \bar{z}'$ for $\Delta_{2,4,6,8}$ by $i\to i{+}1$. Now we can discuss how to perform $\tau$ integrals involving square roots.

\paragraph*{The $\tau$ integrals
}
To perform these $\tau$ integrals, it is convenient to rationalize square roots by a change of variable based on rational points of the quadratic. For example, the rational parameterization $(x,y)=(2t/(1-t^{2}),(1+t^{2})/(1-t^{2}))$ of the quadratic $y^{2}=x^{2}+1$ is obtained by inserting $y-1=(x-0)t$. Let's consider the collinear limit of the four-mass box coefficient $f_{2,4,6,9}$ as an example, which has $\tau=\infty$ as its rational point, and it becomes a rational function of $t$ under the substitution $\tau=\frac{\rho(t-z)(t-\bar{z})}{(t-\sigma)}$ where we define
\begin{equation*}
        \rho=\frac{\langle 1238\rangle\langle 1256\rangle\langle3478\rangle}{\langle2378\rangle\langle1(34)(56)(82)\rangle} \:,\:
        \sigma=\frac{\langle 1348\rangle\langle3456\rangle\langle1278\rangle}{\langle 3478\rangle\langle1(34)(56)(82)\rangle} \:,
\end{equation*}
with the notation $\langle a(bc)(de)(fg)\rangle=\langle abde\rangle\langle acfg\rangle-\langle acde\rangle\langle abfg\rangle$. Now the square root in $z$ enters the final result via the limit of integration $
\int_{0}^{\infty} \mathrm{d} \tau \to \int_{z}^{\infty} \mathrm{d} t \: \frac{\mathrm{d}\tau}{\mathrm{d}t}$.
Similarly, one can find change of variables for the other 3 four-mass boxes, {\it e.g.} the square root in $z'$ comes from the box $f_{3,5,7,9}$. Interestingly, the other two boxes $f_{2,4,7,9}$ and $f_{2,5,7,9}$, which have both $\tau=0$ \emph{and} $\tau=\infty$ as rational points, do not contain algebraic letters after the substitution. This argument explains why there is no square roots in NMHV heptagons from N$^2$MHV octagons. 

A crucial check at this stage is the convergence of $\tau$ integrals. First we collect all terms proportional to $\log\epsilon$ and numerically verify that they integrate to zero. Next, each term of $\tau$-integrand naively has poles at $\tau=0$ or $\infty$, we numerically verified that they are spurious poles and can be removed by a simple subtraction. Once this is done we can easily perform all the subtracted $\tau$ integrals {\it e.g.} using~{\sc PolyLogTools}~\cite{Duhr:2019tlz}.

\paragraph*{Last entry conditions and DCI}

Even before performing the $\tau$ integral, we have found 147 linear independent objects of the form  $Y_{8,1}\bar{Q}\log\frac{\langle\bar{n}i\rangle}{\langle\bar{n}j\rangle}$, from performing $\operatorname{Res}_{\epsilon=0}\int \epsilon\mathrm{d}\epsilon\int \mathrm{d}^{3}\chi_{9}$ on all $Y_{9,2}$'s; out of them, only 17 elements are of the form $[i\,j\,k\,l\,m]\, \bar{Q}\log\frac{\langle\bar{n}(i\,j)\cap(k\,l\,m)\rangle}{\langle\bar{n} i\rangle\langle j k l m\rangle}$, while others do not involve intersections of momentum twistors. After taking cyclic permutations, we have a basis of $8\times 147=1176$ such objects, which is an all-loop prediction for NMHV octagon. The last step is converting these final entries to DCI combinations to remove ambiguities from the kernel of $\bar Q$.
To do this, we transform the above basis to a basis with minimal number of elements that break DCI by adding 0 of the form (\ref{qbarker}). Asking their coefficients to vanish gives constraints on our weight-3 functions from $\tau$ integrals. 
There are $21$ constraints, each being lengthy combination of weight-3 Goncharov polylogarithms, and we have numerically checked that they all vanish. This leaves us with $1155$ DCI combinations as last entry conditions. 

\section{\label{sec:3.5} Results and Consistency Checks}

The differential of two-loop NMHV octagon can be written as $\mathrm{d} R_{8,1}^{(2)}=\sum_\alpha ([i\,j\,k\,l\,m]~\mathrm{d}\log x)_\alpha ~F_\alpha$, where for each combination of $R$ invariant times $\mathrm{d}\log$ of DCI $x$, we have a function $F_\alpha$; these weight-$3$ Goncharov polylogarithms are recorded in {\tt funcdata.m}. From here it is trivial to obtain the complete symbol, ${\mathcal S} [R_{8,1}^{(2)}]$; one can expand the symbol in a basis of ${7 \choose 4}=35$ $R$ invariants, {\it e.g.} $[i\,j\,k\,l\,8]$ for $1\leq i<j<k<l<8$, and we record symbols of the $35$ coefficients, ${\cal S}_{i,j,k,l}$ in {\tt symdata.m}. On average, each symbol has around $10^5$ terms.

The symbol alphabet of $R_{8,1}^{(2)}$ consists of 180 multiplicative-independent rational letters, which can form $180-8=172$ DCI cross-ratios, and 18 independent algebraic letters which we choose to write in a DCI form directly. The 180 rational letters are contained in the prediction of rational symbol alphabet from Landau equations~\cite{Prlina:2017tvx}, but we find that the last class of them is missing. For completeness we list them again:
\begin{itemize}
        \item  68 Pl\"{u}cker coordinates of the form $\langle a\,a{+}1\,b\,c\rangle$,
        \item 8 cyclic images of $\langle 12 \bar{4}\cap \bar{7}\rangle$,
        \item 40 cyclic images of $\langle 1(23)(45)(78)\rangle$, $\langle 1(23)(56)(78)\rangle$, $\langle 1(28)(34)(56)\rangle$, $\langle 1(28)(34)(67)\rangle$, $\langle 1(28)(45)(67)\rangle$,
        \item 48 dihedral images of $\langle 1(23)(45)(67)\rangle$, $\langle 1(23)(45)(68)\rangle$, $\langle 1(28)(34)(57)\rangle$,
        \item 8 cyclic images of $\langle \bar{2}\cap(245)\cap\bar{8}\cap (856)\rangle$,
        \item  8 distinct dihedral images of $\langle \bar{2}\cap(245)\cap\bar{6}\cap (681)\rangle$.
\end{itemize}

The 18 algebraic letters represent the first appearance of square roots in a two-loop amplitudes~\footnote{Similar letters with square roots appear in certain two-loop integrals contributing to MHV and NMHV amplitudes~\cite{Bourjaily:2018aeq,Henn:2018cdp}, but it is unclear which of them may get cancelled in the final answer.} and let's spell them out. The collinear integrals over $9$-point four-mass boxes result in $22 \times 2$ algebraic letters of the form
\begin{equation} \label{irrationalletterform}
    \biggl\{ \frac{x_{\ast}-z}{x_{\ast}-\bar{z}}\,, \quad \frac{x'_{\ast}-z'}{x'_{\ast}-\bar{z}'} \biggr\}
\end{equation}
where $z, z'$ {\it etc.} are defined in~\eqref{zzbar}; 4 of these are known: $\frac{z}{\bar{z}}$, $\frac{1{-}z}{1{-}\bar{z}}$ in $f_{1,3,5,7}$ and their cyclic images in $f_{2,4,6,8}$ (with $z'$), which correspond to $x,x'=0,1$ respectively. The $40$ new algebraic letters take the form of \eqref{irrationalletterform} with $x_{\ast}$ and $x_{\ast}'$ generated by cyclic shifts of the following 5 seeds~\footnote{This set of algebraic letters is not invariant under the reflection, but the new letters obtained by reflection depend on these 44 algebraic letters multiplicatively.}:
\begin{gather}\label{xseeds}
        x_{a}= \frac{\langle 1(52)(34)(78)\rangle\langle 3456\rangle}{\langle 1345\rangle\langle 1256\rangle\langle 3478\rangle} \:, \quad x_{b}=x_{a}\vert_{5\leftrightarrow 6}\:,\\
        x_{c}=\frac{\langle 1378\rangle\langle3456\rangle}{\langle1356\rangle \langle3478\rangle} \:,\:\: x_{d}=x_{c}\vert_{3\leftrightarrow4} \:,\: \:x_{e}=\frac{\langle  187(34)\cap(256)\rangle}{\langle1256\rangle\langle 3478\rangle}\nonumber
\end{gather}
where $i\leftrightarrow j$ denotes the exchange of particle labels. These algebraic letters are not independent since they satisfy multiplicative relations such as
\begin{equation*}
    \frac{x_{a}-z}{x_{a}-\bar{z}}\times \left(\frac{x_{a}-z}{x_{a}-\bar{z}}\right)
    _{i\to i+4} =\frac{z}{\bar{z}}\frac{1-\bar z}{1-z}
\end{equation*}
and we end up with only 18 independent letters~\footnote{We choose to write these letters of the form (\ref{irrationalletterform}) because their multiplicative dependence does not involve rational letters; we keep all 44 letters in our result which makes the answer manifestly cyclic.}.

We find that all algebraic letters always enter the symbols of $\{F_\alpha\}$ in the following combinations 
\begin{equation}
        \biggl(u\otimes \frac{1-z}{1-\bar{z}}+v\otimes \frac{\bar{z}}{z}\biggr)\otimes \frac{x_{\ast}-z}{x_{\ast}-\bar{z}}
\end{equation}
and similarly for $x', z'$. In other words, any algebraic letter with non-trivial $x$ appears only in the third entry, which is accompanied by the corresponding (symbol of) four-mass box in the first two entries. 

It is interesting to rewrite the algebraic letters in the more familiar form $\frac 1 2 (a\pm \sqrt{a^2-4 b})$, where we have $44$ pairs of $(a, b)$ that are polynomials of Pl\"{u}cker coordinates. 
We remark that the discriminant $a^2-4 b$ is always proportional to that for one of the two four-mass boxes, $\Delta^2_{1,3,5,7}$ or $\Delta^2_{2,4,6,8}$, which are square-root branch points from Landau equations~\cite{Prlina:2017azl,Prlina:2017tvx}. However, Landau equations can not predict what $a$ and $b$ can appear, except that branch points $b=0$ correspond to zero locus of some rational letters listed above. We have checked that this is indeed the case for all $b$ of the $44$ letters , {\it e.g.} 
\begin{align*}
(x_a-z)(x_a-\bar{z})&\propto \langle 1(34)(56)(78)\rangle \langle 5 (12)(34)(78)\rangle\,,\\
(x_c-z)(x_c-\bar{z})&\propto \langle 1(34)(56)(78)\rangle \langle 3 (12)(56)(78)\rangle\,,\\
(x_e-z)(x_e-\bar{z})&\propto \langle 1(34)(56)(78) \rangle \langle 2 (34)(56)(78)\rangle\,,
\end{align*}
where we dropped some Pl\"{u}cker coordinates. Like rational ones, algebraic letters are positive in positive kinematics. In {\tt demo\_octagon.nb}, we list all $172+18$ DCI letters which are written using a $9$-dimensional positive parametrization for the momentum twistors~\cite{speyer2005tropical,postnikov2006total}.

In the following we present various consistency checks of our results.

\newcommand{\myparag}[1]{\vspace{1ex}\noindent {\bf #1}}

\myparag{Dual conformal invariance and cyclicity.} The dual conformal symmetry is manifest after we convert last entries to DCI, but the cyclicity is not manifest anymore. Nevertheless we have checked that the answer is cyclic. 

\myparag{First entry conditions.} For a physical scattering amplitude, first entries of its symbol only contain physical poles, {\it i.e.} $\langle i\,i{+}1\,j\,j{+}1\rangle$. We have explicitly checked this, which is true only after converting last entries. 

\myparag{Integrability.} It is crucial that the coefficient of the differential for each basis $R$ invariant, which is of the form $ \sum_i F_i d\log x_i$, can be integrated to a function. This is guaranteed if $\sum_i \mathrm{d}\log F_i\wedge \mathrm{d}\log x_i=0$. In terms of the symbol, the result is integrable if and only if the symbol vanishes after taking last two entries to be the wedge of their $\mathrm{d}\log$'s  (see \cite{Chen:1977oja,brown2009multiple}). We numerically check this using the positive parametrization mentioned above: for each $R$ invariant, all weight-2 symbols which are coefficients of $\binom 92 = 36$ linearly independent $2$-forms vanish.

\myparag{Collinear limits.} We check that the NMHV octagon reduces to NMHV and MHV heptagon upon taking the $k$-preserving and $k$-decreasing collinear limits respectively. We consider the limit $8||7$ by sending 
\begin{equation*}
    Z_8\to Z_7+
      \epsilon \frac{\langle 1257\rangle }{\langle 1256\rangle } Z_6 + 
      \epsilon \tau \frac{\langle 2567\rangle }{\langle 1256\rangle } Z_1 + \eta \frac{\langle 1567\rangle }{\langle 1256\rangle } Z_2,
\end{equation*}
for fixed $\tau$ then taking the limit $\eta\to 0$ before $\epsilon\to 0$. Under the $k$ preserving limit, $R$ invariants behave as  $[a b c 7 8]\to 0$ and $[a b c d 8] \to [abcd7]$, while under the $k$ decreasing limit, the $R$ invariants behave as $[12a78]\to1$ with the others vanishing. After taking such limits and keeping leading terms of $\eta$ and $\epsilon$, it is highly non-trivial that the limits do not depend on the parameters $\eta$, $\epsilon$ and $\tau$, {\it i.e.} it has smooth limits. Moreover, we find the results precisely given by two-loop MHV and NMHV heptagons, which can be found in \cite{CaronHuot:2011ky,Dixon:2016nkn} respectively. 

\myparag{Cancellation of spurious poles.} Finally we check that the residue on any spurious pole of $R$ invariants vanishes. For instance, $\langle 1235\rangle=0$ is a non-physical pole of $[1 2 3 5 8]$, thus the residue as $\langle 1235\rangle \to 0$ should vanish. By cyclicity, we only need to check the following poles
\[
\langle 1235\rangle, \langle 1236\rangle, \langle 1237\rangle, \langle 1246\rangle, 
 \langle 1247\rangle, \langle 1257\rangle, \langle 1357\rangle,
\]
each of which is of the form $\langle 1abc\rangle$ and belongs to exactly one $R$ invariant in our basis. The cancellation of the pole $\langle 1abc\rangle$ means that the coefficient of the corresponding $R$ invariant vanishes as $\langle 1abc\rangle \to 0$. To see this, we send $Z_{1} \to \alpha Z_{a}+\beta Z_{b}+\gamma Z_{c} +\delta Z_{8}$
for fixed $\alpha,\beta,\gamma$, and verified numerically that the coefficient of $[1abc8]$ vanishes under the limit of $\delta \to 0$.
It is curious to note that the cancellation of spurious poles holds for rational and algebraic part separately.

\section{\label{sec:5} Conclusion and Discussions}

Using $\bar{Q}$ equations based on Yangian symmetry of planar ${\cal N}=4$ SYM, we computed the differential of the two-loop NMHV octagon, whose symbol contains interesting algebraic letters. We emphasize that there is no qualitative difference for $n>8$: first we have computed the universal final entries, $Y_{n,1} \times d\log (a)$ to all $n$; furthermore, all algebraic letters and weight-3 functions involving them are computed from four-mass boxes in the same we as we did for $n=8$. We are particularly interested in the explicit result for $n=9$, which via $\bar{Q}$ equations can produce the three-loop MHV octagon~\cite{toapp}. We expect it to be the first three-loop amplitudes with algebraic letters, and it is highly desirable to see how the symbol alphabet changes comparing to two-loop octagons. It is plausible that with sufficient knowledge of symbol alphabet one can initiate the bootstrap for {\it octagon function}, which can push the frontier to higher loops and reveal more interesting structures. 

Relatedly, it is of great interests to further explore mathematical structures of our results. For rational letters, it should be within reach to obtain the symbol alphabet for all $n$ at two loops. The connections to cluster algebra, tropical Grassmannian {\it etc.}~\cite{Cachazo:2019ngv,*Drummond:2019qjk} are certainly worth exploring. It is conceivable that our algebraic letters can be interpreted from tropical $G_+(4,8)$ and the corresponding cluster algebra. A remarkable property called {\it cluster adjacency}~\cite{Drummond:2017ssj} and the closely-related {\it extended Steinmann relations}~\cite{Caron-Huot:2018dsv,*Caron-Huot:2019bsq}, have been observed for hexagons and heptagons, as well as conjectured for higher $n$~\cite{Drummond:2018dfd,*Golden:2018gtk,*Golden:2019kks,*Mago:2019waa}. We have verified that in the list of all-loop final entries to all $n$, at least those involving intersections (which are not affected by subtraction) nicely satisfy this property; we leave the study of adjacency and extended Steinmann relations for the full symbol to future works. 
 
In addition to providing new data points for amplitudes, it would be fascinating to investigate potential interplay of $\bar{Q}$ equations with Feynman integral computations (see \cite{Chicherin:2018ubl,*Chicherin:2018rpz} for recent works along this line). As a preliminary example, we find that the component for two-loop NMHV octagon, $\chi_1 \chi_3 \chi_5 \chi_7$, is completely free of algebraic letters. In our basis it is given by the coefficient of $[13578]$, which is the simplest one with $28376$ terms and $76$ rational letters only, as shown in the ancillary file. 
Our prediction is verified by a computation of certain double-pentagon integrals contributing to this component~\footnote{J. Bourjaily, private communications. In their computation, the integrals and the component in turn are given at a special kinematic point; evaluating our symbol at that point gives the same result.}.
It is obvious from $\bar{Q}$ equations that this component receives no contributions from the $\int \mathrm{d}^{2\vert 3}\mathcal{Z}_{9}$ integrals of four-mass boxes $f_{2,4,6,9}$ and $f_{3,5,7,9}$ and cyclic copies of such integrals, since each of these integrals yield $R$ invariants that involve six adjacent particles only.

\paragraph*{Acknowledgments}
We thank N. Arkani-Hamed, J. Bourjaily, S. Caron-Huot, J. Drummond, \"{O}mer G\"{u}rdo\u{g}an, A. McLoed and M. Spradlin for stimulating discussions and comments on our results. S.H. thanks Center for Mathematical Sciences and Applications, Harvard, for hospitality during the completion of the work. This work is supported in part by the Thousand Young Talents program, the Key Research Program of Frontier Sciences of CAS under Grant No. QYZDBSSW-SYS014, Peng Huanwu center under Grant No. 11747601 and National Natural Science Foundation of China under Grant No. 11935013.




\bibliography{references}

\end{document}